\documentclass[sigplan,nonacm]{acmart}

\AtBeginDocument{
  }

\setcopyright{cc}
\acmYear{2025}

\def\DefaultCutFileName{\def\CommentCutFile{\jobname.cut}}
\DefaultCutFileName

\usepackage{xcolor}
\definecolor{LightGray}{gray}{0.95}

\usepackage[frozencache=true,cachedir=minted-cache,newfloat]{minted} 
\setminted[]{autogobble, breaklines, tabsize=2, fontsize=\footnotesize}
\newminted{dafny}{autogobble, breaklines, tabsize=2, linenos, bgcolor=LightGray}
\newminted{sml}{autogobble, breaklines, tabsize=2, linenos, bgcolor=LightGray}
\newcommand{\inlinedafny}[1]{\mintinline{dafny}{#1}}
\newcommand{\inlinesml}[1]{\mintinline[escapeinside=||]{sml}{#1}}

\usepackage{graphicx}

\begin{document}

\title{Baking for Dafny: A CakeML Backend for Dafny}

\author{Daniel Nezamabadi}
\orcid{0009-0006-8590-435X}
\affiliation{
  \institution{Chalmers University of Technology and University of Gothenburg}
  \city{Gothenburg}
  \country{Sweden}
}
\additionalaffiliation{
  \institution{ETH Zurich}
  \city{Zurich}
  \country{Switzerland}
}
\email{n.daniel@mailbox.org}

\author{Magnus Myreen}
\orcid{0000-0002-9504-4107}
\affiliation{
  \institution{Chalmers University of Technology and University of Gothenburg}
  \city{Gothenburg}
  \country{Sweden}
}
\email{myreen@chalmers.se}

\begin{abstract}
  Dafny is a verification-aware programming language that allows
  developers to formally specify their programs and prove them
  correct. Currently, a Dafny program is compiled in two steps: First,
  a backend translates the input program to a high-level target
  language like C\# or Rust. Second, the translated program is
  compiled using the target language's toolchain. Recently, an
  intermediate representation (IR) has been added to Dafny that serves
  as input to new backends. At the time of writing, none of these
  steps are verified, resulting in both the backend and the target
  language's toolchain being part of Dafny's trusted computing base
  (TCB). To reduce Dafny's TCB, we started developing a new backend
  that translates Dafny to CakeML, a verified, bootstrapped subset of
  Standard ML, in the interactive theorem prover HOL4. We also started
  to define functional big-step semantics for the Dafny IR to prove
  correctness of the backend.
\end{abstract}



\maketitle

\section{Introduction}
Dafny \cite{dafny} is a verification-aware programming language which
gives developers the ability to provide specifications and prove that
their programs are correct with respect to them. Dafny programs are
compiled in two steps: First, the Dafny program is translated into
some target language like C\# or Rust via a backend. Second, the
translated program is compiled using the target language's
toolchain. Recently, an intermediate step has been added to the first
step, which allows backends to translate an intermediate
representation (IR) instead. This is how the Rust backend is
implemented.

Currently, both the backend used in the first step, and the toolchain
used in the second are trusted to preserve the semantics of the original
program and thus are part of Dafny's trusted computing base (TCB). This
represents a serious risk, given that any bug in these components might
cause the compiled program to behave unexpectedly. Indeed, recent work
on testing Dafny's toolchain has discovered a number of compiler
bugs \cite{dafny-test, dafny-test-2}.

To reduce Dafny's TCB, we started developing a new backend for Dafny in the
interactive theorem prover HOL4 \cite{hol4-overview}\footnote{The
development of the backend can be found at \url{https://github.com/CakeML/cakeml/tree/master/compiler/dafny}.}.
The backend translates Dafny to CakeML \cite{cakeml-overview}, a verified, bootstrapped implementation of a subset of
Standard ML.

Once a Dafny program has been translated to CakeML, the verified
nature of the CakeML compiler guarantees that its semantics are
preserved down to the generated binary \cite{cakeml-verified}.
By defining the bulk of the backend
in HOL4, we can define functional big-step semantics
\cite{functional-big-step} for the Dafny IR and prove that the backend
preserves them for any valid Dafny IR program. To extract an
executable backend out of our HOL4 definitions, we use CakeML's
translator, which generates an equivalent CakeML program, and a proof showing
its semantic equivalence to the original definitions \cite{cakeml-translator}.

\section{Implementation}
This section gives an overview of the steps taken to compile
from Dafny to CakeML.

\subsection{Overview}
\begin{figure}
  \centering
  \includegraphics[trim=57 97 55 97, clip, width=\columnwidth]{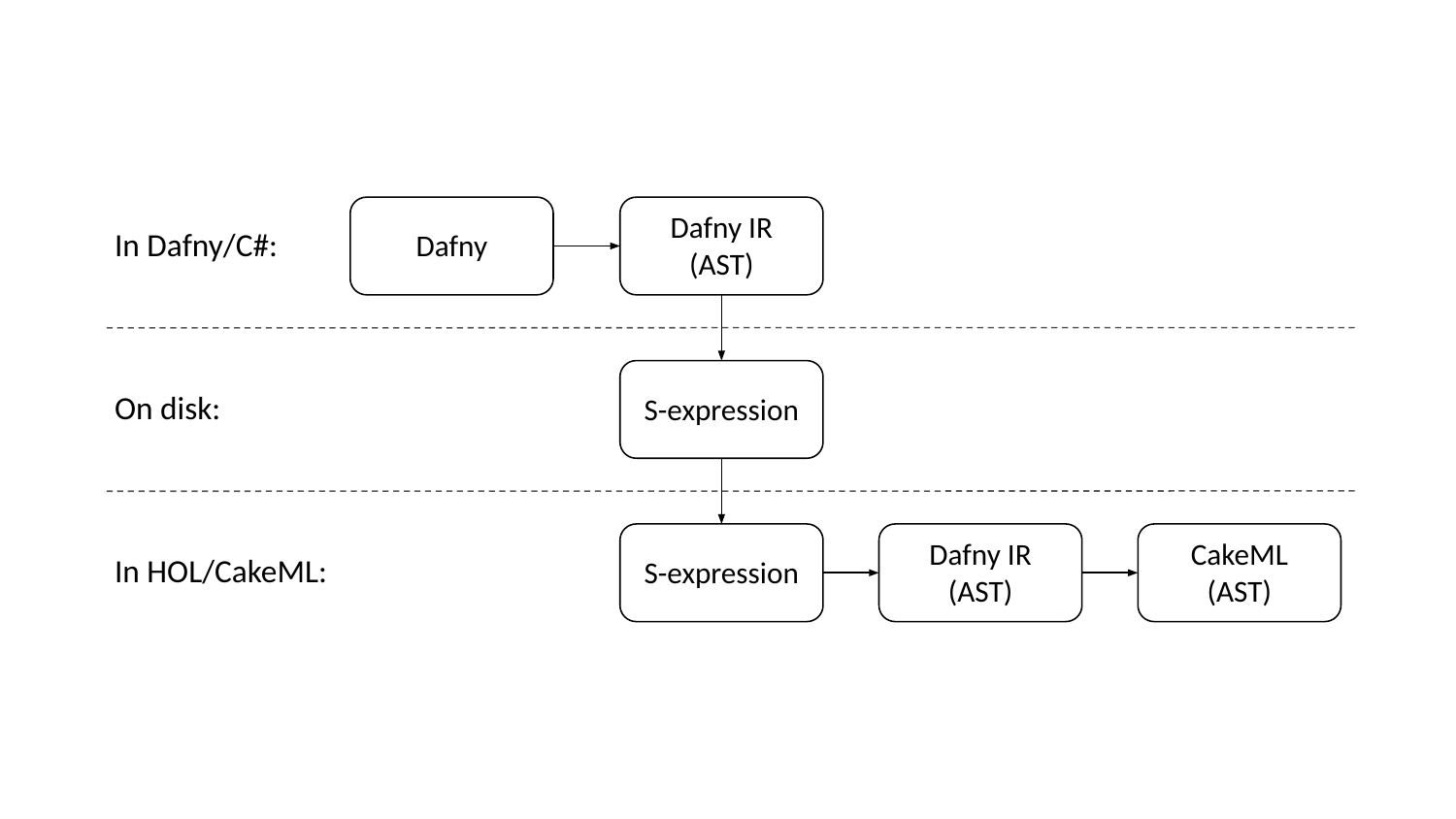}
  \caption{Transformation steps from a Dafny program to the CakeML
    AST.}
  \Description{Fully described in the text.}
  \label{fig:imp:overview}
\end{figure}

As visualized in Figure~\ref{fig:imp:overview}, the Dafny program is
first translated into its intermediate representation. This step is
part of Dafny, and mostly implemented in C\#. However, the abstract
syntax tree (AST) of the intermediate representation is defined in
Dafny. Note that this step is part of the TCB.

As motivated in the introduction, we want to implement the majority of
the backend in HOL4/CakeML. Thus, we ``export'' the
intermediate representation of the program by writing it as an
S-expression to a file.

The file containing the S-expression is then read, lexed and
parsed in HOL4. The resulting S-expression value is then ``parsed'' to
recover the abstract syntax tree of the intermediate
representation.

Finally, the Dafny IR AST is compiled into a CakeML AST. The main
challenge is to fit a modern, multi-paradigm programming language into
CakeML, which is an SML-style functional programming language. For
example, while functions in CakeML cannot refer to functions that are
defined later in textual order, Dafny does not have such
limitations. We deal with this problem by marking all functions within
a module to be mutually recursive in CakeML, allowing them to refer to
each other freely. In the long term, we plan to use a dependency
graph to untangle non-mutually recursive functions.

\subsection{Examples}
We will now go through a series of Dafny programs and explain how
features of the Dafny language can be compiled to CakeML.

\paragraph{Hello World}
Consider Figure~\ref{code:dfy:hello-world}, which contains a single
\verb|Main| method and prints ``Hello, Cake''. It is compiled into
the CakeML code given in Figure~\ref{code:cml:hello-world}. Note that
the compiler actually generates an AST -- the representations in
Figure~\ref{code:cml:hello-world} and \ref{code:cml:loop} have been
created by massaging the output of the CakeML pretty-printer.

As can be seen in line 1 of Figure~\ref{code:cml:hello-world},
\verb|Main| has been placed into the default module called
\verb|_module|. This is a direct translation of the program's
representation in the Dafny IR, which defines a program as a list of
modules.

Dafny methods are compiled into CakeML functions that return
\verb|Unit|. We use exceptions to deal with the fact that a method can
return at any point. In general, return statements are compiled by
raising an exception, with an exception handler being installed around
the method body. \inlinesml{Dafny.Return} is defined as
\inlinesml{exception Return} in a separate module called \verb|Dafny|,
which contains what other backends refer to as the Dafny runtime.

Dafny functions have to be compiled in the same way, as the concept
of a function currently does not exist on the Dafny IR level. Instead,
both methods and functions become a list of statements. It may be
interesting to investigate whether it is possible to encode at least a
subset of functions as expressions, as this may lead to improved
performance for backends that target functional languages, including CakeML.

We represent strings as lists of characters in CakeML. Another
possibility would be to use the \verb|string| type, which is a packed
vector of characters and may be more efficient. We chose to use the
former representation, as it maps cleanly to Dafny's \verb|string|
type, which is a synonym for a sequence of characters. As CakeML's
\inlinesml{print} requires values of type \inlinesml{string}, we need
to add a call to \inlinesml{String.implode} to convert from a list
of characters to a string.

\begin{figure}
\begin{dafnycode}
  method Main() {
    print "Hello, Cake\n";
  }
\end{dafnycode}
\caption{``Hello World'' in Dafny.}\label{code:dfy:hello-world}
\Description{Fully described in the text.}
\end{figure}

\begin{figure}
\begin{smlcode}
structure _module = struct
  fun Main =
    (print (String.implode ("Hello, Cake\n");
     raise Dafny.Return)
    handle Dafny.Return => ();
end
\end{smlcode}
\caption{Compilation of ``Hello World'' to CakeML. We do not
  include the definition of the Dafny module for
  brevity.} \label{code:cml:hello-world}
\Description{Fully described in the text.}
\end{figure}

\paragraph{Factorial}
We now consider an imperative implementation of the factorial
function as shown in Figure~\ref{code:dfy:factorial}.
\begin{figure}
\begin{dafnycode}
  method Factorial(n: int) returns (result: int) {
    result := 1;
    var i := 1;
    while i <= n {
      result := result * i;
      i := i + 1;
    }
  }
\end{dafnycode}
\caption{Loop-based factorial function in Dafny.}\label{code:dfy:factorial}
\Description{Fully described in the caption.}
\end{figure}

\begin{figure}
\begin{minted}[autogobble, breaklines, tabsize=2, linenos, bgcolor=LightGray, escapeinside=||]{sml}
let fun CML_while_0 =
  if not ((! n) < (! _|0|_i)) then 
    (result := (! result) * (! _|0|_i);
     _|0|_i := (! _|0|_i) + 1;
     CML_while_0 ())
  else ()
in
  (CML_while_0 ()) handle Dafny.Break => ()
\end{minted}
\caption{Compilation of the loop in Figure~\ref{code:dfy:factorial} to
  CakeML. We do not include the declaration and initialization of
  variables for brevity.}\label{code:cml:loop}
\Description{Fully described in the text.}
\end{figure}

As variables in Dafny are mutable, we compile them as references. For example,
\inlinedafny{var i := 1} is compiled to \inlinesml{let val _|0|_i = ref 0 in
  _|0|_i := 1}. Since CakeML integers have arbitrary precision by default, we
can directly map Dafny integers onto CakeML integers. Note that \verb|i| is set
to 1 after the initialization of the reference. This mirrors the behavior in the
Dafny IR, which splits declaration and initialization in this case. Also note
that CakeML requires references to be always initialized with a value. While
this is straightforward for simple types like integers, booleans and strings, it
is less obvious for data types or type variables. We defer this discussion to
Section~\ref{sec:state}.

CakeML does not define any while-like constructs. Instead, we compile
Dafny loops into tail-recursive functions in CakeML and use exceptions
for (early) return statements as shown in Figure~\ref{code:cml:loop},
which is a modification of the (derived) \verb|while| form defined in
Standard ML \cite{sml-def}. On entry, we first check whether the loop
condition holds or not. Since variables are already encoded as
references, no extra bookkeeping is necessary to keep track of values
between the recursive function calls. Like return statements, break
statements are encoded using exceptions. The main difference is that
the exception handler is not installed around the loop body, but the
first invocation of the recursive function. Note that compiling Dafny
loops into tail-recursive functions does not cause stack overflows,
as tail calls are compiled to jumps rather than calls, and thus
do not add to the stack depth.

Note that in Figure~\ref{code:cml:loop}, there is an exception handler
even though no exceptions are actually raised. This is due to the
original loop not containing any \verb|break| or \verb|continue|
statements, and the compiler not doing any kind of analysis at the
moment to detect this case.

We also use exceptions to model labeled
statements, breaking to a label, and \verb|continue|. In these cases,
the exception contains a string with the label name, and exceptions
are handled as shown in Figure~\ref{code:cml:label}.

\begin{figure}
\begin{smlcode}
(body) handle Dafny.LabeledBreak l =>
  if l = lbl then ()
  else raise (Dafny.LabeledBreak l)
\end{smlcode}
\caption{Handling labeled breaks in CakeML. \texttt{lbl} is the label for \texttt{body}.}\label{code:cml:label}
\Description{Fully described in the text.}
\end{figure}

\section{Correctness Statement}\label{sec:cor}
\begin{figure}
  \centering
  \includegraphics[trim=172 117 172 126, clip, width=\columnwidth]{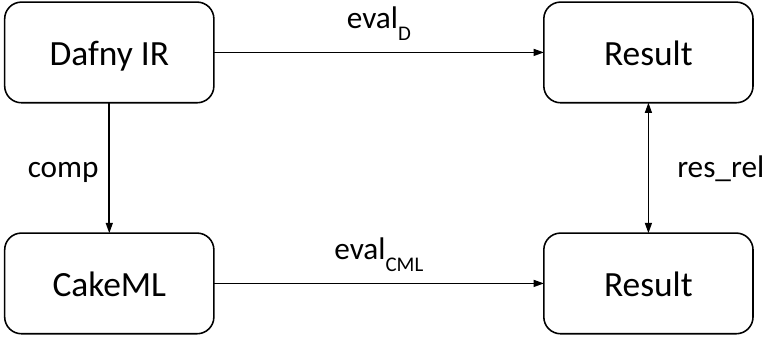}
  \caption{Overview of the simulation theorem. We do not visualize
    the state for brevity.}
  \label{fig:imp:proof-overview}
\end{figure}

To prove that the compiler preserves the semantics of the program in
its intermediate representation, we must first define the semantics of
the Dafny IR. We achieve this by defining functional big-step
semantics \cite{functional-big-step} for the Dafny IR, which can be
thought of as an interpreter with a decreasing clock on recursive
calls to ensure termination. Once the semantics have been defined, the
correctness statement can be summarized as ``Assuming equivalent
initial states, the result of evaluating the CakeML translation of the
Dafny IR program should be the same as directly evaluating the Dafny
IR program''. We formalize this statement in
a simulation theorem~\eqref{eq:correct}, where $p$ represents the
Dafny IR program, $s$, $s'$, and $r$ the
states and result of the Dafny IR evaluator,
and $t$, $t'$, $r'$ the states and result of the CakeML evaluator.
Note that we only consider valid programs, that is,
programs that terminate without running into type errors. A visualization can be found in Figure~\ref{fig:imp:proof-overview}.
\begin{align} \label{eq:correct}
  \forall s \, p \, s' \, r \, t. \, & \text{eval}_{\text{D}} \, s \, p = (s', r) \wedge r \neq \text{Err} \wedge \text{state\_rel} \, s \, t \nonumber\\
  & \Rightarrow \exists t' \, r'. \, \text{eval}_{\text{CML}} \, t \, (\text{comp} \, p) = (t', r') \, \wedge \\
  & \phantom{\Rightarrow \exists t' \, r' . \,} \text{state\_rel} \, s' \, t' \wedge \text{res\_rel} \, r \, r' \nonumber
\end{align}

At the time of writing, we have only proven a similar statement for the compilation
of some binary expressions.

\section{Future Plans}\label{sec:state}
Our current plan is to extend the compiler in both the constructs it can
compile, and its correctness proofs.

In particular, we expect that data types, type parameters, traits, and
classes are going to be the most interesting constructs to support.

For the former two features, recall that we model variables as references, which
must be initialized in CakeML. This poses a challenge for non-trivial types like
user-defined data types or type parameters. For data types, it may be possible
to make use of the ``grounding constructor'' in Dafny, which provides an
instance of a data type. For type parameters, one could imagine using the fact
that Dafny enforces definite assignment rules if a type cannot be automatically
initialized, by initializing the corresponding references with dummy values
which are guaranteed to be overwritten later in the program. This would require
us to disable CakeML's static type checker.  Note that as a consequence of the
simulation theorem described in Section~\ref{sec:cor}, this does not influence
overall soundness. However, as the static type checker has been a useful tool to
catch bugs during development, we would like to use it for as long as possible.

One could imagine using monomorphization to keep the static type
checker enabled. However, this comes with its own problems: Dafny
allows the definition of types that instantiate their type variables in
the recursive application, e.g., \inlinedafny{datatype Foo<A> = Nil |
  Pair(Foo<(A, A)>)}. Thus, naively trying to monomorphize this
datatype for integers would require us to generate
\inlinedafny{Foo<int>}, \inlinedafny{Foo<(int, int)>},
\inlinedafny{Foo<((int, int), (int, int))>} and so on.

In the case of traits and classes, we believe that it will be necessary to
represent objects using tuples and something like virtual function tables, as
CakeML does not support objects natively. Similar to type parameters, this will
most likely require us to disable CakeML's static type checker, as classes may
have a wider interface than the traits they may implement, resulting in tuples
of different sizes.

In the long term, our goal is to upstream the CakeML backend into the
Dafny repository. In this regard, investigating ways to make the
compiler more maintainable and efficient is going to be
important. For the latter, measuring the impact of encoding functions
as expressions on the Dafny IR level, or the impact of using lists of
characters instead of strings, may be possible first steps.

To further reduce the TCB, future work could involve defining semantics
for Dafny, and verifying that the translation to the Dafny IR maintains
the original program semantics. This would essentially correspond to
applying our current approach to Dafny's frontend. Alternatively,
one could imagine directly translating Dafny to CakeML, skipping the
Dafny IR. We note, however, that future backends that translate the
Dafny IR would not be able to benefit from this approach.

\begin{acks}
  Many thanks to João C. Pereira, Fabio Madge, and the anonymous
  reviewers for their feedback on this extended abstract, and
  Clément Pit-Claudel for insightful discussions.
  This work was funded by an Amazon Research Award.
\end{acks}

\bibliographystyle{ACM-Reference-Format}
\bibliography{bibfile}


\begin{thebibliography}{9}


\ifx \showCODEN    \undefined \def \showCODEN     #1{\unskip}     \fi
\ifx \showDOI      \undefined \def \showDOI       #1{#1}\fi
\ifx \showISBNx    \undefined \def \showISBNx     #1{\unskip}     \fi
\ifx \showISBNxiii \undefined \def \showISBNxiii  #1{\unskip}     \fi
\ifx \showISSN     \undefined \def \showISSN      #1{\unskip}     \fi
\ifx \showLCCN     \undefined \def \showLCCN      #1{\unskip}     \fi
\ifx \shownote     \undefined \def \shownote      #1{#1}          \fi
\ifx \showarticletitle \undefined \def \showarticletitle #1{#1}   \fi
\ifx \showURL      \undefined \def \showURL       {\relax}        \fi
\providecommand\bibfield[2]{#2}
\providecommand\bibinfo[2]{#2}
\providecommand\natexlab[1]{#1}
\providecommand\showeprint[2][]{arXiv:#2}

\bibitem[Donaldson et~al\mbox{.}(2024)]%
        {dafny-test-2}
\bibfield{author}{\bibinfo{person}{Alastair~F. Donaldson},
  \bibinfo{person}{Dilan Sheth}, \bibinfo{person}{Jean-Baptiste Tristan}, {and}
  \bibinfo{person}{Alex Usher}.} \bibinfo{year}{2024}\natexlab{}.
\newblock \showarticletitle{Randomised Testing of the Compiler for a
  Verification-Aware Programming Language}. In \bibinfo{booktitle}{\emph{2024
  IEEE Conference on Software Testing, Verification and Validation (ICST)}}.
  \bibinfo{pages}{407--418}.
\newblock
\urldef\tempurl%
\url{https://doi.org/10.1109/ICST60714.2024.00044}
\showDOI{\tempurl}


\bibitem[Irfan et~al\mbox{.}(2022)]%
        {dafny-test}
\bibfield{author}{\bibinfo{person}{Ahmed Irfan}, \bibinfo{person}{Sorawee
  Porncharoenwase}, \bibinfo{person}{Zvonimir Rakamari\'{c}},
  \bibinfo{person}{Neha Rungta}, {and} \bibinfo{person}{Emina Torlak}.}
  \bibinfo{year}{2022}\natexlab{}.
\newblock \showarticletitle{Testing Dafny (experience paper)}. In
  \bibinfo{booktitle}{\emph{Proceedings of the 31st ACM SIGSOFT International
  Symposium on Software Testing and Analysis}} (Virtual, South Korea)
  \emph{(\bibinfo{series}{ISSTA 2022})}. \bibinfo{publisher}{Association for
  Computing Machinery}, \bibinfo{address}{New York, NY, USA},
  \bibinfo{pages}{556–567}.
\newblock
\showISBNx{9781450393799}
\urldef\tempurl%
\url{https://doi.org/10.1145/3533767.3534382}
\showDOI{\tempurl}


\bibitem[Kumar et~al\mbox{.}(2014)]%
        {cakeml-overview}
\bibfield{author}{\bibinfo{person}{Ramana Kumar}, \bibinfo{person}{Magnus~O.
  Myreen}, \bibinfo{person}{Michael Norrish}, {and} \bibinfo{person}{Scott
  Owens}.} \bibinfo{year}{2014}\natexlab{}.
\newblock \showarticletitle{{CakeML}: A Verified Implementation of {ML}}. In
  \bibinfo{booktitle}{\emph{Principles of Programming Languages ({POPL})}}.
  \bibinfo{publisher}{ACM Press}, \bibinfo{pages}{179--191}.
\newblock
\urldef\tempurl%
\url{https://doi.org/10.1145/2535838.2535841}
\showDOI{\tempurl}


\bibitem[Leino(2010)]%
        {dafny}
\bibfield{author}{\bibinfo{person}{K.~Rustan~M. Leino}.}
  \bibinfo{year}{2010}\natexlab{}.
\newblock \showarticletitle{Dafny: an automatic program verifier for functional
  correctness}. In \bibinfo{booktitle}{\emph{Proceedings of the 16th
  International Conference on Logic for Programming, Artificial Intelligence,
  and Reasoning}} (Dakar, Senegal) \emph{(\bibinfo{series}{LPAR'10})}.
  \bibinfo{publisher}{Springer-Verlag}, \bibinfo{address}{Berlin, Heidelberg},
  \bibinfo{pages}{348–370}.
\newblock
\showISBNx{3642175104}


\bibitem[Milner et~al\mbox{.}(1997)]%
        {sml-def}
\bibfield{author}{\bibinfo{person}{Robin Milner}, \bibinfo{person}{Robert
  Harper}, \bibinfo{person}{David MacQueen}, {and} \bibinfo{person}{Mads
  Tofte}.} \bibinfo{year}{1997}\natexlab{}.
\newblock \bibinfo{booktitle}{\emph{{The Definition of Standard ML}}}.
\newblock \bibinfo{publisher}{The MIT Press}.
\newblock
\showISBNx{9780262287005}
\urldef\tempurl%
\url{https://doi.org/10.7551/mitpress/2319.001.0001}
\showDOI{\tempurl}


\bibitem[Myreen and Owens(2014)]%
        {cakeml-translator}
\bibfield{author}{\bibinfo{person}{Magnus~O. Myreen} {and}
  \bibinfo{person}{Scott Owens}.} \bibinfo{year}{2014}\natexlab{}.
\newblock \showarticletitle{Proof-producing Translation of Higher-order logic
  into Pure and Stateful {ML}}.
\newblock \bibinfo{journal}{\emph{Journal of Functional Programming ({JFP})}}
  \bibinfo{volume}{24}, \bibinfo{number}{2-3} (\bibinfo{date}{May}
  \bibinfo{year}{2014}), \bibinfo{pages}{284--315}.
\newblock
\urldef\tempurl%
\url{https://doi.org/10.1017/S0956796813000282}
\showDOI{\tempurl}


\bibitem[Owens et~al\mbox{.}(2016)]%
        {functional-big-step}
\bibfield{author}{\bibinfo{person}{Scott Owens}, \bibinfo{person}{Magnus~O.
  Myreen}, \bibinfo{person}{Ramana Kumar}, {and} \bibinfo{person}{Yong~Kiam
  Tan}.} \bibinfo{year}{2016}\natexlab{}.
\newblock \showarticletitle{Functional Big-step Semantics}. In
  \bibinfo{booktitle}{\emph{European Symposium on Programming ({ESOP})}}
  \emph{(\bibinfo{series}{Lecture Notes in Computer Science},
  Vol.~\bibinfo{volume}{9632})}, \bibfield{editor}{\bibinfo{person}{Peter
  Thiemann}} (Ed.). \bibinfo{publisher}{Springer}, \bibinfo{pages}{589--615}.
\newblock
\urldef\tempurl%
\url{https://doi.org/10.1007/978-3-662-49498-1_23}
\showDOI{\tempurl}


\bibitem[Slind and Norrish(2008)]%
        {hol4-overview}
\bibfield{author}{\bibinfo{person}{Konrad Slind} {and} \bibinfo{person}{Michael
  Norrish}.} \bibinfo{year}{2008}\natexlab{}.
\newblock \showarticletitle{A Brief Overview of HOL4}. In
  \bibinfo{booktitle}{\emph{Theorem Proving in Higher Order Logics}},
  \bibfield{editor}{\bibinfo{person}{Otmane~Ait Mohamed},
  \bibinfo{person}{C{\'e}sar Mu{\~{n}}oz}, {and} \bibinfo{person}{Sofi{\`e}ne
  Tahar}} (Eds.). \bibinfo{publisher}{Springer Berlin Heidelberg},
  \bibinfo{address}{Berlin, Heidelberg}, \bibinfo{pages}{28--32}.
\newblock
\showISBNx{978-3-540-71067-7}


\bibitem[Tan et~al\mbox{.}(2019)]%
        {cakeml-verified}
\bibfield{author}{\bibinfo{person}{Yong~Kiam Tan}, \bibinfo{person}{Magnus~O.
  Myreen}, \bibinfo{person}{Ramana Kumar}, \bibinfo{person}{Anthony Fox},
  \bibinfo{person}{Scott Owens}, {and} \bibinfo{person}{Michael Norrish}.}
  \bibinfo{year}{2019}\natexlab{}.
\newblock \showarticletitle{The verified {CakeML} compiler backend}.
\newblock \bibinfo{journal}{\emph{Journal of Functional Programming}}
  \bibinfo{volume}{29} (\bibinfo{year}{2019}).
\newblock
\urldef\tempurl%
\url{https://doi.org/10.1017/S0956796818000229}
\showDOI{\tempurl}


\end{thebibliography}

\end{document}